\documentclass[aps,prb,twocolumn]{revtex4}
\usepackage{graphicx}
\usepackage{bm}

\begin{document}

\small

\title{Coulomb blockade of non-local electron transport in metallic conductors}

\author{D.S. Golubev}
\affiliation{Institut f\"ur Nanotechnologie,
Karlsruher Institut f\"ur Technologie (KIT), 76021 Karlsruhe, Germany}
\author{A.D. Zaikin}
\affiliation{Institut f\"ur Nanotechnologie,
Karlsruher Institut f\"ur Technologie (KIT), 76021 Karlsruhe, Germany}
\affiliation{I.E. Tamm Department of
Theoretical Physics, P.N. Lebedev Physics Institute, 119991
Moscow, Russia}


\begin{abstract}
We consider a metallic wire coupled to two metallic electrodes
via two junctions placed nearby. A bias voltage applied to one of such junctions alters the electron
distribution function in the wire in the vicinity of another junction thus modifying both its noise and the Coulomb blockade correction to its conductance. We evaluate such interaction corrections to both local
and non-local conductances demonstrating non-trivial
Coulomb anomalies in the system under consideration.
Experiments on non-local electron transport with Coulomb effects can be
conveniently used to test inelastic electron relaxation in metallic conductors at low temperatures.

\end{abstract}

\maketitle

\section{Introduction}
\label{sec:introduction}

A direct relation between shot noise and Coulomb
blockade of electron transport in mesoscopic conductors is well known. In normal conductors this
relation was established theoretically  \cite{GZ2001,yeyati} and subsequently 
confirmed experimentally \cite{pierre}.
Later the same ideas were extended to subgap electron transport in
normal-superconducting (NS) hybrids \cite{GZ09}. The latter results appear to
provide an adequate interpretation for experimental observations \cite{Bezr} of Coulomb
effects in such systems.

While all the above developments concern local electron transport and
shot noise, the question arises if there also exists any general relation between 
non-locally correlated shot noise in multi-terminal conductors and Coulomb 
effects on non-local electron transport in such systems. An important
example is provided by three-terminal NSN structures which have recently 
received a great deal of attention in both experiments
\cite{Beckmann,Teun,Venkat,Basel} and theory \cite{thycar} in connection with
the phenomenon of crossed Andreev reflection. The latter phenomenon yields
non-trivial behavior of the non-local subgap conductance in such
structures. Further interesting features emerge if one takes into account
electron-electron interactions. One can observe, for example, the sign change
of the non-local conductance caused either by the influence of the
electromagnetic modes propagating along the wire \cite{levy_yeyati_nature}, or
by positive cross-correlations in non-local current noise
\cite{an_refl_coul}. Furthermore, positive cross-correlations in shot noise
are directly linked to Coulomb ani-blockade of non-local electron transport
\cite{an_refl_coul,levy_yeyati_prb}. Thus, a general relation between
cross-correlated shot noise and Coulomb effects in non-local subgap electron
transport in NSN systems turns out to be much richer
than that in the local case \cite{GZ09}. 

\begin{figure}
\begin{center}
\includegraphics[width=7.5cm]{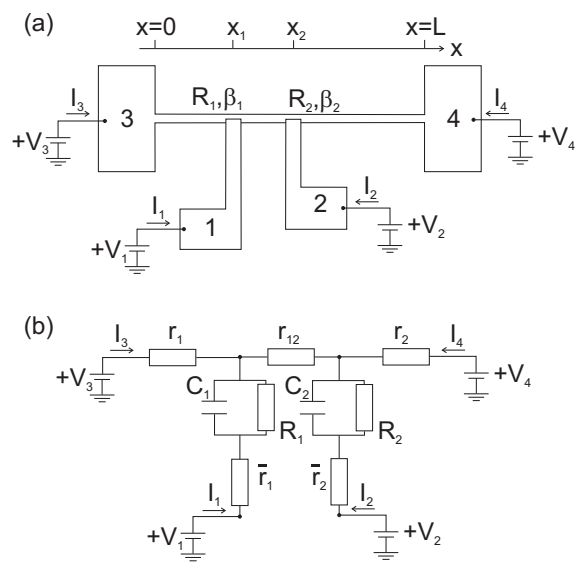}
\end{center}
\caption{(a) Schematics of the system under consideration. It consists of two metallic
electrodes 1 and 2 coupled to a metallic wire of length $L$ connecting the electrodes 3 and 4 via the two junctions with resistances $R_1,R_2$, Fano factors $\beta_1,\beta_2$ and capacitances $C_1,C_2$.
(b) Equivalent electric circuit of the system depicted in panel (a).
}
\label{fig_two_junctions}
\end{figure}

In this paper we will address the impact of
electron-electron interactions on non-local effects in normal metallic 
structures depicted in Fig. \ref{fig_two_junctions}.
Non-local properties of such systems turn out to be very sensitive to 
inelastic processes. At low temperatures such processes in metallic
conductors usually become rather weak and electrons can propagate at long distances, typically of order microns, without suffering any significant energy changes. Hence, provided voltage bias is applied to a mesoscopic conductor, its electron distribution function $f(E)$ may substantially  deviate from its equilibrium value universally defined by the Fermi function $f_F(E)=1/(1+e^{E/T})$. For example, low temperature distribution function $f(E)$ may take the characteristic double-step form in comparatively short metallic wires attached to two big reservoirs with different electrostatic potentials \cite{saclay}.

Further interesting effects emerge if one takes into account an interplay
between non-equilibrium effects and electron-electron interactions. 
Consider, e.g., a tunnel junction between two metallic leads.
Provided the the junction resistance  
significantly exceeds that of the leads,
the effect of Coulomb interaction can be modeled by introducing
interactions between electrons and some linear electro-magnetic environment
\cite{SZ,Ingold}. In this case the strength of Coulomb interaction is
characterized by an effective impedance of the environment and the current
across the tunnel junction reads 
\begin{eqnarray}
I(V) &=& \frac{1}{eR}\int dE_L dE_R
\nonumber\\ &&\times\,
\big\{ f_L(E_L)\big[1-f_R(E_R)\big] P(E_L-E_R-eV)
\nonumber\\ &&
-  \big[1-f_L(E_L)\big]f_R(E_R)P(-E_L+E_R+eV) \big\},
\label{PofE}
\end{eqnarray}
where $f_{L,R}(E)$ are the electron distribution functions in the left and
right electrodes and $P(E)$ is the probability to excite a photon with energy $E$ due to interaction between the junction and the environment. Provided the environment has a non-zero impedance and both distribution functions $f_L(E)$ and $f_R(E)$ are close to the Fermi function,
Eq. (\ref{PofE}) yields the well known zero-bias anomaly on the I-V curve, i.e. the Coulomb blockade dip in the differential conductance $dI/dV$ in the limit of low voltages \cite{PZ,SZ,Ingold}.  Furthermore, should at least one of the distribution functions deviate from the equilibrium one, the I-V curve can receive further significant modifications. For instance,
if one distribution function takes the double step form \cite{saclay}, it
follows immediately from Eq. (\ref{PofE}) that the Coulomb blockade dip in the
conductance should split into two separate dips . These dips can be -- and
have been \cite{Anthore} -- detected experimentally thus offering a
possibility to investigate non-equilibrium effects with the aid of small
capacitance tunnel junctions as it was demonstrated, e.g., by experimental
analysis of the impact of magnetic impurities on 
inelastic relaxation of electrons in normal metals \cite{Anthore,Huard}.

Despite clear advantages and simplicity of Eq. (\ref{PofE}), it might not always be convenient to employ in order to analyze combined effects
of non-equilibrium and Coulomb interaction in metallic conductors.
Indeed, the applicability of Eq. (\ref{PofE}) is restricted
to junctions with very low barrier transmissions, i.e. the effect of higher
transmissions cannot be correctly accounted for my means of this equation. The
latter effect might be important, in particular if one needs to evaluate the
non-local conductance. In addition, the function $P(E)$ is usually evaluated
under the assumption of thermodynamic equilibrium in electromagnetic
environment, 
which effectively implies equilibrium electron distributions in both leads. If, however,
the electron subsystem is driven out of equilibrium, self-consistent
evaluation of $P(E)$ 
might become a non-trivial problem. Furthermore,
the function $P(E)$ would in general be difficult to evaluate for effectively 
non-linear electromagnetic environments.

The above complications are avoided within the kinetic equation analysis presented
below. This approach only requires resistances of metallic leads to remain
smaller than the quantum resistance unit $h/e^2$. Within the same theoretical
framework it allows to evaluate both non-local shot noise and the effect of
electron-electron interactions on non-local electron transport in normal
metallic conductors as well as to describe a non-trivial interplay between Coulomb
effects and inelastic processes in such structures.

The paper in organized as follows. In Sec. \ref{sec:model} we outline our model and define the Hamiltonian of our system. In Sec. \ref{sec:noise}
we analyze non-local correlated shot noise in the system under consideration.
In Sec. \ref{sec:current}  we extend this analysis taking into account electron-electron interactions
and demonstrating direct relation between shot noise and interaction effects in non-local electron transport. A brief summary of our key observations is contained in Sec. \ref{sec:summary}.
Some technical details are relegated to Appendices. In Appendix \ref{sec:derivation} we outline key steps of our derivation of the kinetic equation employed in our analysis. Necessary details of our solution of this kinetic equation are displayed in Appendix \ref{sec:solution}.

\section{The model}
\label{sec:model}

In this paper we will consider the system depicted in Fig. \ref{fig_two_junctions}. It consists of a metallic wire of length $L$ connected to two leads
1 and 2 by two small area junctions located at $\bm{x}=\bm{x}_j=(x_j,0,0)$, $j=1,2$  and two
bulk reservoirs 3 and 4 at $x=0$ and $x=L$ ($x$ is the coordinate along the wire).

The system depicted in Fig. \ref{fig_two_junctions} is described by the Hamiltonian
\begin{eqnarray}
H=H_1+H_2+H_{\rm wire}+H_{T,1}+H_{T,2},
\end{eqnarray}
where
\begin{eqnarray}
H_j=\sum_{\alpha=\uparrow,\downarrow}\int d{\bm x}\,
\hat \psi^\dagger_{j,\alpha}({\bm x})\left(-\frac{\nabla^2}{2m}-\mu\right)\hat \psi_{j,\alpha}({\bm x}), \;\; j=1,2,
\nonumber
\end{eqnarray}
are the Hamiltonians of the normal metals, 
\begin{eqnarray}
H_{\rm wire} &=& \sum_{\alpha=\uparrow,\downarrow}\int d{\bm x}\,
\hat\chi^\dagger_{\alpha}({\bm x})\bigg(-\frac{\nabla^2}{2m}-\mu
\nonumber\\ &&
+\,U(\bm{x}) +eV(t,\bm{x})\bigg)\hat\chi_{\alpha}({\bm x})
\end{eqnarray}
is the Hamiltonian of the wire and
\begin{eqnarray}
H_{T,j}=\sum_{\alpha}\int_{{\cal A}_j} d^2{\bm x}
\big[ t_j({\bm x})\, e^{i\phi_j(t)}\,  \hat\psi^\dagger_{j,\alpha}({\bm x}) \hat\chi_{\alpha}({\bm x}) + {\rm c.c.}  \big]
\label{HT}
\end{eqnarray}
are tunneling Hamiltonians describing transfer of electrons across the contacts with area ${\cal A}_j$ and tunneling amplitude $t_j({\bm r})$. Here and below $m$ stands for the electron mass, $\mu$ is the chemical potential, the index $\alpha$ labels the spin projection, the potential $U(\bm{x})$ accounts for disorder inside the wire and $V(t,{\bm x})$ represents the scalar potential. 
The transmissions of the conducting channels of the junctions are related to the matrix elements of the tunnel amplitudes $t_n^{(j)}$
between the states belonging to the same conducting channel as follows
\begin{equation}
T^{(j)}_n=\left|\tau_n^{(j)}\right|^2={4\pi^2\nu_j\nu_0\left|t_n^{(j)}\right|^2}/{\left(1+\pi^2\nu_j\nu_0\left|t_n^{(j)}\right|^2\right)^2},
\label{ttr}
\end{equation}
where $\nu_j$ ($j=1,2$) is the density of states in the corresponding terminal and $\nu_0$ is the density of states inside the wire. The barrier resistances $R_1$ and $R_2$ and
their Fano factors $\beta_1$ and $\beta_2$ are expressed in a standard way as
\begin{eqnarray}
\frac{1}{R_j}= \frac{2e^2}{h}\sum_n T_n^{(j)}, \; \beta_j={\sum_n T_n^{(j)}\left(1-T_n^{(j)}\right)}\bigg/{\sum_n T_n^{(j)}}.
\end{eqnarray}
A voltage bias, respectively $V_1,V_2,V_3$ and $V_4$, can be applied to all four metallic terminals 1,2,3 and 4.

In the setup of Fig. \ref{fig_two_junctions} one of the junctions, e.g. the junction 2, may be viewed as an injector, which drives electron distribution function in the wire out of equilibrium.
The junction 1 may then be used as a detector for experimental investigation of nonequilibrium effects. One of the ways to observe such effects is to study the non-local differential conductance $\partial I_1/\partial V_2$ of our system. Clearly, in such kind of experiments the distance between the junctions
should not exceed an effective electron inelastic relaxation length $L_{\rm in}(T)$ which sets the scale for non-equilibrium effects in the wire at a given temperature.
Thus, the setup of Fig. \ref{fig_two_junctions} may be used to directly measure $L_{\rm in}$.

Finally we note that the above particular system geometry is chosen merely for the sake of definiteness. The key steps of our subsequent analysis and the results obtained from it remain applicable to a much broader class of systems than that depicted in Fig. \ref{fig_two_junctions}.
E.g., the wire may be replaced by a metallic lead of any shape, and ultimately all geometry specific details can be absorbed in few elements of the conductance matrix.

\section{Cross-correlated shot noise}
\label{sec:noise}

We begin with the analysis of shot noise employing the so-called Boltzmann-Langevin technique \cite{Sukhorukov,Blanter}
based on a kinetic equation for the electron distribution function $f(t,E,\bm{x})$.
Low frequency cross-correlated shot noise in multi-terminal metallic structures has
already been studied before, see, e.g., Ref. \onlinecite{Sukhorukov}.
Here we will briefly rederive and somewhat extend the corresponding results in order to illustrate the basic idea of the approach in a relatively simple case. In the next section we will extend this approach in order to include electron-electron interactions where more involved calculations will be necessary.

The Boltzmann-Langevin kinetic equation accounts for current noise produced by the junctions 1 and 2 and has the form
\begin{eqnarray}
\frac{\partial f}{\partial t}-D\nabla_{\bm{x}}^2f &=&
-\frac{f-f_F(E-eV(t,\bm{x}))}{\tau_{\rm in}}
\nonumber\\ &&
-\, \frac{f-f_F(E-ew_1)}{2e^2\nu_0R_1}\delta(\bm{x}-\bm{x}_1)
\nonumber\\ &&
-\,\frac{f-f_F(E-ew_2)}{2e^2\nu_0R_2}\delta(\bm{x}-\bm{x}_2)
\nonumber\\ &&
+\,  \frac{\eta_1(t,E)\delta(\bm{x}-\bm{x}_1)+\eta_2(t,E)\delta(\bm{x}-\bm{x}_2)}{2e\nu_0}.
\label{kinetic_0}
\end{eqnarray}
Here $D$ and $\nu_0$ are respectively the electron diffusion constant and
the electron density of states at the Fermi energy inside the wire. We also introduced electrostatic potentials of the leads  $w_1$ and $w_2$
in the vicinity of the junctions 1 and 2,
\begin{eqnarray}
w_j=\left(1-\frac{\bar r_j}{R_j}\right)V_j,\;\;
\bar r_1,\bar r_2\ll R_1,R_2,
\label{w1w2}
\end{eqnarray}
where the resistances of the leads $\bar r_j$ are defined in Fig. \ref{fig_two_junctions}b,
and $\tau_{\rm in}=D/L^2_{\rm in}$ in the inelastic relaxation time.
Note that here we are not going to discuss physical mechanisms dominating the process of electron energy
relaxation at low temperatures and simply treat $\tau_{\rm in}$ as a phenomenological parameter.

The potential $V(t,\bm{x})$ should be determined self-consistently from the equation
\begin{eqnarray}
\int dE\big[ f(t,E,\bm{x})-f_F(E-eV(t,\bm{x}))\big]=0,
\label{self-consist}
\end{eqnarray}
which directly follows from the charge neutrality condition inside the normal metal. 
This charge neutrality condition in metals is a direct consequence of strong Coulomb
interaction between electrons as well as between electrons and lattice ions.
Integrating Eq. (\ref{kinetic_0}) over energy we obtain
\begin{eqnarray}
&& \left(\frac{\partial }{\partial t}-D\nabla_{\bm{x}}^2\right) V(t,\bm{x})=
\nonumber\\ &&
+\,\frac{(w_1-V(t,\bm{x}_1))}{2e^2\nu_0R_1}\delta(\bm{x}-\bm{x}_1)
+\frac{(w_2-V(t,\bm{x}_2))}{2e^2\nu_0R_2}\delta(\bm{x}-\bm{x}_2)
\nonumber\\ &&
+\,\int dE \frac{\eta_1(t,E)\delta(\bm{x}-\bm{x}_1)+\eta_2(t,E)\delta(\bm{x}-\bm{x}_2)}{2e^2\nu_0}.
\label{diffusion_V}
\end{eqnarray}
Note that inelastic relaxation time $\tau_{\rm in}$ drops out from this equation.

The stochastic variables $\eta_1(t,E)$ and $\eta_2(t,E)$ in Eqs. (\ref{kinetic_0}) and (\ref{diffusion_V})
account for low frequency fluctuations of the current carried by electrons with
energy $E$ through the junctions 1 and 2 respectively. The corresponding correlators read\cite{Blanter}
\begin{eqnarray}
&& \langle\eta_i(t_1,E_1)\eta_j(t_2,E_2)\rangle = \frac{1}{R_j}\delta_{ij}\delta(t_1-t_2)\delta(E_1-E_2)
\nonumber\\ &&\times\,
\big\{ \beta_jf(t_1,E_1,\bm{x}_j)\left[1-f_F(E_1-ew_j)\right]
\nonumber\\&&
+\, \beta_j\left[1-f(t_1,E_1,\bm{x}_j)\right]f_F(E_1-ew_j)
\nonumber\\&&
+\,(1-\beta_j)f(t_1,E_1,\bm{x}_j)\left[1-f(t_1,E_1,\bm{x}_j)\right]
\nonumber\\&&
+\,(1-\beta_j)f_F(E_1-ew_j)\left[1-f_F(E_1-ew_j)\right]
\big\}.
\label{corr_eta_0}
\end{eqnarray}
Finally, no fluctuations occur
at fully open contacts between the wire and the terminals 3 and 4. These contacts are
accounted for by the boundary conditions
\begin{equation}
f(t,E,x=0)=f_F(E-eV_3),\; f(t,E,x=L)=f_F(E-eV_4).
\label{bou}
\end{equation}
Note that in the Eq. (\ref{kinetic_0}) we have neglected the internal current noise generated in the wire \cite{Sukhorukov}.
In order to justify this approximation, in what follows we will assume
\begin{eqnarray}
r_1,r_2,r_{12},\bar r_1,\bar r_2\ll R_1,R_2,
\label{resistances}
\end{eqnarray}
i.e. we will assume the junction resistances to be much higher than the resistances of the metallic leads and the wire
(see Fig. \ref{fig_two_junctions}b for the definition of the resistances).
Thus, the task at hand is to solve Eqs. (\ref{kinetic_0}), (\ref{diffusion_V}) supplemented by Eqs.  (\ref{corr_eta_0}), (\ref{bou}) and to evaluate the current noise in our system.

As we already discussed above, the form of the distribution function inside the wire may essentially depend on the relation
between its size $L$ and the inelastic relaxation length $L_{\rm in}$.
Yet another relevant parameter to be compared with $L_{\rm in}$ is the distance between the two junctions $|x_2-x_1|$.
Provided inelastic relaxation is very strong, $L_{\rm in} \ll |x_2-x_1| < L$, the inelastic term in Eq. (\ref{kinetic_0})
plays the dominant role and the electron distribution function $f$ in the wire remains close to the Fermi function $f_F(E-eV(\bm{x}))$
with the voltage $V(\bm{x})$ to be derived from Eq. (\ref{diffusion_V}).
In the opposite weak relaxation limit $L \ll L_{\rm in}$
the inelastic collision integral in Eq. (\ref{kinetic_0}) can be neglected.
Of interest is also the intermediate limit of a long wire $L \gg L_{\rm in}$ but relatively
weak relaxation $|x_2-x_1| \ll L_{\rm in}$.

We begin our analysis by defining the currents $I_1$ and $I_2$
across junctions 1 and 2:
\begin{eqnarray}
I_j(t) = \frac{1}{eR_j}\int dE \left[ f_F(E-ew_j)-f(t,E,\bm{x}_j)\right]+\delta\tilde I_j,
\label{current_0}
\end{eqnarray}
where
$
\delta\tilde I_j = \int dE\,\eta_j(t,E)
$
is the fluctuating current in the $j$-th junction.
In the limit of full inelastic relaxation, $L_{\rm in}\ll |x_2-x_1|<L$, the distribution function in the wire
has the equilibrium form, and
with the aid of Eq. (\ref{corr_eta_0}) we derive the zero frequency spectral noise power $\tilde S_j = \int dt \langle\delta\tilde I_j(t)\delta\tilde I_j(0)\rangle$,
\begin{eqnarray}
\tilde S_j = \beta_j\frac{ev_j}{R_j}\coth\frac{ev_j}{2T}+(1-\beta_j)2T,
\label{S_bare}
\end{eqnarray}
where $v_j=w_j-V(\bm{x}_j)$ are voltage drops across the junctions. Under the condition (\ref{resistances}) one finds
\begin{eqnarray}
&& v_j = \left(1-\frac{\bar r_j}{R_j}\right)V_j - \frac{r_2V_3+r_1V_4}{r_1+r_2},\;\; j=1,2.
\label{v1v2}
\end{eqnarray}
Naturally, Eq. (\ref{S_bare}) just coincides with the noise power for a perfectly voltage biased junction \cite{Blanter}.

\begin{figure}
\includegraphics[width=8cm]{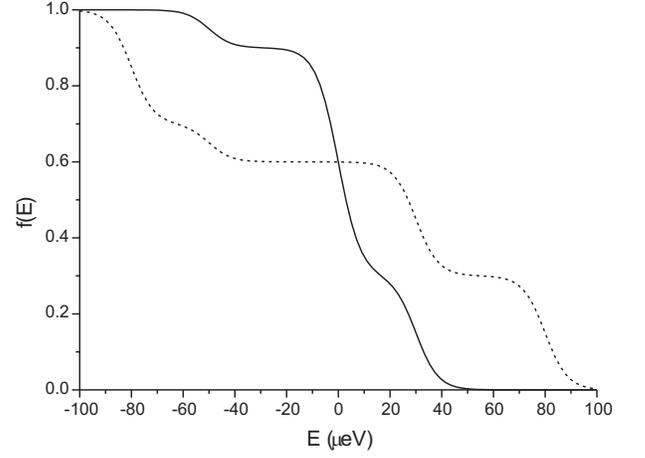}
\caption{Electron distribution function in the wire in the vicinity of the first junction, $f(E,\bm{x}_1)$.
The solid line shows $f(E,\bm{x}_1)$ given by Eq. (\ref{distribution_2}) which is applicable at intermediate values of inelastic relaxation. In this regime the distribution function has three steps. The dashed line corresponds to elastic limit in which the distribution function $f(E,\bm{x}_1)$ is defined in Eq. (\ref{distribution_elastic}). The system parameters are: $T=50$ mK, ${\tilde{\cal D}(0,\bm{x}_1,\bm{x}_1)}/{2e^2\nu_0R_1}=0.3$,
${\tilde{\cal D}(0,\bm{x}_1,\bm{x}_2)}/{2e^2\nu_0R_2}=0.1$, $w_1=30$ $\mu$V, $w_2=-50$ $\mu$V, $V_3=80$ $\mu$V, $V_4=-80$ $\mu$V,
$r_1/r=r_2/r=0.5$.}
\label{fig_fofE}
\end{figure}

Let us now consider the limit $|x_2-x_1|<L_{\rm in}\ll L$. In this case,
according to Eq. (\ref{kinetic_0}) the electron distribution function $f(t,E,\bm{x}_j)$ deviates from the equilibrium form and fluctuates. Hence,
the total current noise should acquire an additional contribution. In order to proceed let us establish the relation
between the distribution functions $f(t,E,\bm{x}_j)$ and the stochastic variables $\eta_j$. This goal can be achieved
with the aid of the diffuson ${\cal D}(t,\bm{x},\bm{x}')$, which is defined as a solution of the diffusion equation
\begin{eqnarray}
\left(\frac{\partial}{\partial t}-D\nabla_{\bm{x}}^2
+ \frac{\delta(\bm{x}-\bm{x}_1)}{2e^2\nu_0R_1} + \frac{\delta(\bm{x}-\bm{x}_2)}{2e^2\nu_0R_2}\right){\cal D}(t,\bm{x},\bm{x}')=
\nonumber\\
-\frac{1}{\tau_{\rm in}}{\cal D}(t,\bm{x},\bm{x}')+\delta(t)\delta(\bm{x}-\bm{x}')
\label{diffusion}
\end{eqnarray}
with boundary conditions
\begin{eqnarray}
{\cal D}(t,0,\bm{x})={\cal D}(t,\bm{x},0)={\cal D}(t,L,\bm{x})={\cal D}(t,\bm{x},L)=0.
\end{eqnarray}
The physical meaning of the diffuson ${\cal D}(t,\bm{x},\bm{x}')$ is well known: It defines the probability for an electron injected into the wire at the point $\bm{x}'$ to
reach the point $\bm{x}$ during the time $t$.
We also define the Fourier transformed diffuson
$$
\tilde{\cal D}(\omega,\bm{x},\bm{x}')=\int dt\, e^{i\omega t}{\cal D}(t,\bm{x},\bm{x}').
$$
The solution of Eq. (\ref{kinetic_0}) can be expressed in the form
\begin{eqnarray}
&& f(t,E,\bm{x}) =\int dt' d^3\bm{x}' \frac{{\cal D}(t-t',\bm{x},\bm{x}')}{\tau_{\rm in}}f_F\left(E-eV(t',\bm{x}')\right)
\nonumber\\ &&
+\,\frac{\tilde{\cal D}(0,\bm{x},\bm{x}_1)}{2e^2\nu_0R_1}f_F(E-ew_1)
+\frac{\tilde{\cal D}(0,\bm{x},\bm{x}_2)}{2e^2\nu_0R_2}f_F(E-ew_2)
\nonumber\\ &&
+\,\frac{1}{2e\nu_0}\int dt'\big[{\cal D}(t-t',\bm{x},\bm{x}_1)\eta_1(t',E)
\nonumber\\ &&
+\,{\cal D}(t-t',\bm{x},\bm{x}_2)\eta_2(t',E)\big].
\label{distribution_1}
\end{eqnarray}
This general expression gets simplified in the limit $e|V_3-V_4|\ll TL/L_{\rm in}$
and provided current fluctuations can be neglected, i.e. $\eta_{1,2}\to 0$. In this case the electric potential $V(\bm{x})$ does not depend on time and slowly varies in space. Then one can approximately replace $f_F\left(E-eV(t',\bm{x}')\right)$
by $f_F\left(E-eV(\bm{x})\right)$. Afterwards, employing the properties of the diffuson, one finds
\begin{eqnarray}
&& f(E,\bm{x}) =\left[1-\frac{\tilde{\cal D}(0,\bm{x},\bm{x}_1)}{2e^2\nu_0R_1}-\frac{\tilde{\cal D}(0,\bm{x},\bm{x}_2)}{2e^2\nu_0R_2}\right]f_F\left(E-eV(\bm{x})\right)
\nonumber\\ &&
+\,\frac{\tilde{\cal D}(0,\bm{x},\bm{x}_1)}{2e^2\nu_0R_1}f_F(E-ew_1)
+\frac{\tilde{\cal D}(0,\bm{x},\bm{x}_2)}{2e^2\nu_0R_2}f_F(E-ew_2).
\label{distribution_2}
\end{eqnarray}
The non-equilibrium distribution function in this regime has three steps, see also Fig. \ref{fig_fofE}.
The first one comes from the distribution function of the isolated wire $f_F\left(E-eV(\bm{x})\right)$, while the
other two steps, $\propto f_F(E-ew_j)$, originate from the junctions. Since the diffuson $\tilde{\cal D}(0,\bm{x},\bm{x}')$
decays at distances $|\bm{x}-\bm{x}'|>L_{\rm in}$, the distribution function acquires its equilibrium form far away from the junctions.

The currents $I_1$ and $I_2$ can be evaluated with the aid of Eqs. (\ref{self-consist}) and (\ref{diffusion_V}). They read
\begin{eqnarray}
I_1 &=& G_{11}v_1 - G_{\rm nl} v_2 + \delta I_1,
\nonumber\\
I_2 &=& - G_{\rm nl} v_1 + G_{22}v_2 + \delta I_2.
\label{current}
\end{eqnarray}
Here $G_{jj}$ and $G_{\rm nl}$ define respectively local and non-local conductances of our structure:
\begin{eqnarray}
G_{jj} = \frac{1}{R_j}-\frac{\tilde{\cal D}_0(0,\bm{x}_j,\bm{x}_j)}{2e^2\nu_0R_j^2},\;\;
G_{\rm nl} = \frac{\tilde{\cal D}_0(0,\bm{x}_1,\bm{x}_2)}{2e^2\nu_0R_1R_2},
\label{Gij}
\end{eqnarray}
where $\tilde{\cal D}_0(\omega,\bm{x},\bm{x}')$ is the solution of the diffusion equation (\ref{diffusion}) with $\tau_{\rm in}\to\infty$.
One can equivalently write these conductances in the from
\begin{eqnarray}
G_{jj}=\frac{1}{R_j}-\frac{r_1r_2}{r_1+r_2},\;\;
G_{\rm nl}=\frac{r_1r_2}{(r_1+r_2)R_1R_2}.
\label{Gij_r}
\end{eqnarray}
Here we again assumed that $r_{12}\ll r_1,r_2,$ and $r_1,r_2,\bar r_1,\bar r_2\ll R_1,R_2$.
Let us emphasize that the results (\ref{Gij}), (\ref{Gij_r}) were derived from Eq. (\ref{diffusion_V}) and, hence,
are not sensitive to inelastic relaxation at all. Besides that, general expressions (\ref{Gij}) are not restricted to the wire
geometry and remain valid for any shape of the leads.

Finally, the noise terms $\delta I_j$ appearing in Eq. (\ref{current}) read
\begin{eqnarray}
 \delta I_1 &=& \int dE dt'
\left[\left(\delta(t-t')-\frac{{\cal D}_0(t-t',\bm{x}_1,\bm{x}_1)}{2e^2\nu_0R_1}\right)\eta_1(t',E)
\right.
\nonumber\\ &&
\left.
-\,\frac{{\cal D}_0(t-t',\bm{x}_1,\bm{x}_2)}{2e^2\nu_0R_1}\eta_2(t',E)\right],
\label{xi1}
\\
\delta I_2 &=& \int dE dt'
\left[-\frac{{\cal D}_0(t-t',\bm{x}_2,\bm{x}_1)}{2e^2\nu_0R_2}\eta_1(t',E)
\right.
\nonumber\\ &&
\left.
+\,\left(\delta(t-t')-\frac{{\cal D}_0(t-t',\bm{x}_2,\bm{x}_2)}{2e^2\nu_0R_2}\right)\eta_2(t',E)\right].
\label{xi2}
\end{eqnarray}
Clearly, they differ from the bare noise terms $\delta \tilde I_j$ since the contributions coming from electrons diffusing from one junction to the other or returning back to the same junction are also taken into account.

We are now in position to evaluate the zero frequency noise power matrix
\begin{equation}
S_{ij}=\int dt\langle \delta I_i(t)\delta I_j(0)\rangle.
\label{npm}
\end{equation}
With the aid of Eqs. (\ref{corr_eta_0}) and (\ref{xi1},\ref{xi2}) we express the noise power for the first junction as
\begin{eqnarray}
S_{11} &=& R_1G_{11}^2
\int dE \big\{ \beta_1 f(E,\bm{x}_1)\left[1-f_F(E-ew_1)\right]
\nonumber\\ &&
+\, \beta_1 \left[1-f(E,\bm{x}_1)\right]f_F(E-ew_1)
\nonumber\\ &&
+\, (1-\beta_1) f(E,\bm{x}_1)\left[1-f(E,\bm{x}_1)\right]
\nonumber\\ &&
+\, (1-\beta_1) \left[1-f_F(E-ew_1)\right]f_F(E-ew_1)
\big\}
\nonumber\\&&
+\, R_2 G_{\rm nl}^2
\int dE \big\{ \beta_2 f(E,\bm{x}_2)\left[1-f_F(E-ew_2)\right]
\nonumber\\ &&
+\, \beta_2\left[1-f(E,\bm{x}_2)\right]f_F(E-ew_2)
\nonumber\\ &&
+\, (1-\beta_2) f(E,\bm{x}_2)\left[1-f(E,\bm{x}_2)\right]
\nonumber\\ &&
+\, (1-\beta_2) \left[1-f_F(E-ew_2)\right]f_F(E-ew_2)
\big\}.
\label{noise11_0}
\end{eqnarray}
Substituting the distribution function (\ref{distribution_2}) into this expression,
assuming $G_{\rm nl}\ll G_{11},G_{22}$, defining the function
\begin{eqnarray}
W(v)=ev\coth\frac{ev}{2T},
\end{eqnarray}
and under the condition (\ref{resistances}), we arrive at the final result
\begin{eqnarray}
S_{11}&=& G_{11} \big[ \beta_1W(v_1)+(1-\beta_1)2T\big]
\nonumber\\ &&
+\,    \beta_1\,\tilde G_{\rm nl}\,W(v_1-v_2)
+ (1-\beta_1)\, G_{\rm nl} W(v_2),
\label{S11_final}
\\
S_{12} &=& -G_{\rm nl} \big[ \beta_1W(v_1)
+\,\beta_2 W(v_2)+(2-\beta_1-\beta_2)2T\big].
\label{S12_final}
\end{eqnarray}
Noise power for the second junction $S_{22}$ is defined by Eq. (\ref{S11_final}) with interchanged indices $1\leftrightarrow 2$.
Here we have introduced the effective non-local conductance
\begin{eqnarray}
\tilde G_{\rm nl}=\frac{\tilde{\cal D}(0,\bm{x}_1,\bm{x}_2)}{2e^2\nu_0R_1R_2},
\end{eqnarray}
which, in contrast to $G_{\rm nl}$, is suppressed by inelastic relaxation.
One has $\tilde G_{\rm nl}\ll G_{\rm nl}$ if the distance between the junctions exceeds $L_{\rm in}$,
and $\tilde G_{\rm nl}=G_{\rm nl}$ if $|x_1-x_2|\ll L_{\rm in}$.

The first line of Eq. (\ref{S11_final}) just coincides with the standard expression for the shot noise of a mesoscopic conductor with the Fano factor $\beta_1$,
while the second and third lines provide the corrections induced in the first junction by the second one. The origin of these corrections is simple:
voltage bias applied to the second junction yields modifications in the electron distribution function in the vicinity of the first junction
(cf. Eq. (\ref{distribution_1})) thus changing its current noise.

Now we turn to the regime of a short wire, $L\ll L_{\rm in}$, where inelastic relaxation can be fully ignored.
Accordingly in Eq. (\ref{kinetic_0}) we set $\tau_{\rm in}=\infty$ and repeat the above calculation in this limit. As a result,
the distribution function in the wire acquires the four step shape
\begin{eqnarray}
f(E,\bm{x}) &=& \left[1-\frac{\tilde{\cal D}_0(0,\bm{x},\bm{x}_1)}{2e^2\nu_0R_1}-\frac{\tilde{\cal D}_0(0,\bm{x},\bm{x}_2)}{2e^2\nu_0R_2}\right]
\nonumber\\ &&\times\,
\left[ \frac{r_2}{r}f_F(E-eV_3)+\frac{r_1}{r}f_F(E-eV_4) \right]
\nonumber\\ &&
+\,\frac{\tilde{\cal D}_0(0,\bm{x},\bm{x}_1)}{2e^2\nu_0R_1}f_F(E-ew_1)
\nonumber\\ &&
+\,\frac{\tilde{\cal D}_0(0,\bm{x},\bm{x}_2)}{2e^2\nu_0R_2}f_F(E-ew_2).
\label{distribution_elastic}
\end{eqnarray}
This function is also illustrated in Fig. \ref{fig_fofE}. Here we introduced the total resistance of the wire $r=r_1+r_2+r_{12}$ and assumed $r_{12}\ll r_1,r_2$.
The noise in the limit (\ref{resistances}),
$r_{12}\ll r_1,r_2$ and $\beta_1=\beta_2=1$ becomes
\begin{eqnarray}
S_{11} &=&  G_{11}\bigg[\frac{r_2}{r}W(w_1-V_3) + \frac{r_1}{r}W(w_1-V_4)\bigg]
\nonumber\\ &&
+\,\tilde G_{\rm nl} W(w_1-w_2),
\label{S11_elastic}
\\
S_{12} &=& -G_{\rm nl}\bigg[\frac{r_2}{r}W(w_1-V_3)
+ \frac{r_1}{r}W(w_1-V_4)
\nonumber\\ &&
+\,\frac{r_2}{r}W(w_2-V_3)
+ \frac{r_1}{r}W(w_2-V_4)\bigg].
\label{S12_elastic}
\end{eqnarray}
Comparing these expressions with Eqs. (\ref{S11_final}), (\ref{S12_final}), we observe that
they coincide either provided $V_3=V_4$ or in the large bias limit  $w_j-V_\alpha\gg T$. Otherwise,
every function $W$ entering the result in the limit of strong relaxation splits up into two functions in the limit $L\ll L_{\rm in}$.

\section{Non-local electron transport in the presence of interactions}
\label{sec:current}

Until now we have ignored interaction effects and restricted our consideration to low frequency current fluctuations.
Below we will account for electron-electron interactions and evaluate the interaction correction to the conductance matrix of our system.
Extending the arguments \cite{GZ2001,yeyati}, we will demonstrate
a close relation between Coulomb blockade of non-local electron transport and shot noise in the system under consideration.
For this purpose it will be necessary to go beyond the low frequency limit and allow for arbitrary (not necessarily slow)
fluctuations of voltages $v_j(t)$ across the junctions.
In this regime the time and energy dependent electron distribution function in the wire $f(t,E,\bm{x})$ becomes
ill-defined due to quantum mechanical uncertainty principle. This problem can be cured by employing the Keldysh Green function of electrons
\begin{eqnarray}
G(t_1,t_2,\bm{x})=\int \frac{dE}{2\pi} e^{-iE(t_1-t_2)}\left[1-2f\left(\frac{t_1+t_2}{2},E,\bm{x}\right)\right],
\label{G_Keldysh}
\end{eqnarray}
which fully describes electron dynamics at arbitrarily high frequencies.
Applying the Fourier transformation (\ref{G_Keldysh}) to the kinetic equation (\ref{kinetic_0}) we cast it to the form
\cite{array}
\begin{eqnarray}
&& \left( \frac{\partial}{\partial t_1}+\frac{\partial}{\partial t_2} -D\nabla_{\bm{x}}^2 + \frac{1}{\tau_{\rm in}}
+ i\dot\Phi(t_1,\bm{x}) - i\dot\Phi(t_2,\bm{x})\right)G
\nonumber\\ &&
=\,\frac{1}{\tau_{\rm in}} \frac{-iT e^{-i[\Phi(t_1,\bm{x})-\Phi(t_2,\bm{x})]}}{\sinh\pi T(t_1-t_2)}
\nonumber\\ &&
-\, \frac{\delta(\bm{x}-\bm{x}_1)}{2e^2\nu_0R_1}\left(G-\frac{-iT e^{-i[\phi_1(t_1)-\phi_1(t_2)]}}{\sinh\pi T(t_1-t_2)}\right)
\nonumber\\ &&
-\, \frac{\delta(\bm{x}-\bm{x}_2)}{2e^2\nu_0R_2}\left(G-\frac{-iT e^{-i[\phi_2(t_1)-\phi_2(t_2)]}}{\sinh\pi T(t_1-t_2)}\right)
\nonumber\\ &&
-\, \frac{\delta(\bm{x}-\bm{x}_1)}{e\nu_0}\eta_1(t_1,t_2) - \frac{\delta(\bm{x}-\bm{x}_2)}{e\nu_0}\eta_2(t_1,t_2).
\label{kinetic}
\end{eqnarray}
Here the stochastic variables $\eta_j(t_1,t_2)$, which now also depend on two times, are correlated as follows
\begin{eqnarray}
&&\langle \eta_i(t_1,t_2)\eta_j(t_3,t_4)\rangle =
\nonumber\\ &&
=\, \frac{\delta_{ij}}{8\pi R_j}\bigg[
\frac{2}{\pi^2}\lim_{\epsilon\to 0}\frac{\epsilon^2}{\left((t_1-t_2)+\epsilon^2\right)\left((t_3-t_4)+\epsilon^2\right)}
\nonumber\\ &&
-\, \beta_j\bigg( G(t_1,t_4)\frac{-iT e^{-i[\phi_j(t_3)-\phi_j(t_2)]}}{\sinh\pi T(t_3-t_2)}
\nonumber\\ &&
+\,  \frac{-iT e^{-i[\phi_j(t_1)-\phi_j(t_4)]}}{\sinh\pi T(t_1-t_4)}G(t_3,t_2) \bigg)
\nonumber\\ &&
-\,(1-\beta_j)\bigg( G(t_1,t_4)G(t_3,t_2)
\nonumber\\ &&
+\, \frac{-iT e^{-i[\phi_j(t_1)-\phi_j(t_4)]}}{\sinh\pi T(t_1-t_4)}\frac{-iT e^{-i[\phi_j(t_3)-\phi_j(t_2)]}}{\sinh\pi T(t_3-t_2)} \bigg)
\bigg].
\label{corr_eta}
\end{eqnarray}
In Eqs. (\ref{kinetic}) and (\ref{corr_eta}) we defined the fluctuating phases of the leads $\phi_j=\int_{t_0}^t dt' ew_j(t') $ as well as the phase
$
\Phi(t,\bm{x})=\int_{t_0}^t dt'\, eV(t',\bm{x}),
$
where $V(t',\bm{x})$ is the electric potential inside the wire which fluctuates both in time and in space and includes interaction effects.

Note that fully quantum mechanical description of interaction effects in metallic conductors generally involves two (rather than one) quantum fluctuating phase fields $\Phi_F$ and $\Phi_B$ (defined on the two branches of the Keldysh contour) appearing after the standard Hubbard-Stratonovich decoupling of the Coulomb term in the Hamiltonian \cite{SZ,deph}. Provided interaction effects are sufficiently small (as is the case here, see below) one can effectively eliminate one of these fields, $\Phi_-=\Phi_F-\Phi_B$, and retain only the "center-of-mass" field $\Phi_+=(\Phi_F+\Phi_B)/2 \to \Phi$.
The derivation of the kinetic equation (\ref{kinetic}) in the tunnel limit $\beta_1=\beta_2=1$ is presented in the Appendix \ref{sec:model}.
Rigorous derivation of the kinetic equation (\ref{kinetic}) based on the non-linear $\sigma-$model
as well as its applicability conditions can be found in Ref. \onlinecite{array}.

Now we turn to the expression for the current through the first junction $I_1$.
In order to derive this expression it is necessary to solve the kinetic equation (\ref{kinetic}).
Technical details of this procedure are presented in Appendix \ref{sec:solution}.
Here we directly proceed to the corresponding results.

Let us first consider the limit of strong inelastic relaxation, $L\gg L_{\rm in}$,
and assume that the wire potential varies in space slowly enough, $e|V_3-V_4|\ll TL/L_{\rm in}$.
In this case the current through the first junction acquires the form
\begin{eqnarray}
I_1 &=& G_{11}\left[v_1 - \frac{\beta_1}{e}\int_0^\infty dt \frac{\pi T^2}{\sinh^2\pi Tt} K_{11}(t)\sin[ev_1t]\right]
\nonumber\\ &&
-\, \tilde G_{\rm nl}\frac{\beta_1}{e}\int_0^\infty dt \frac{\pi T^2}{\sinh^2\pi Tt} K_{11}(t) \sin[e(v_1-v_2)t]
\nonumber\\ &&
+\, \frac{\beta_1}{ e}\int dt'dt'' K_{12}(t''-t')\frac{{\cal D}(t',\bm{x}_1,\bm{x}_2)}{2e^2\nu_0R_1R_2}
\nonumber\\ &&\times\,
\frac{\pi T^2}{\sinh^2\pi Tt''}\sin[(v_1-v_2)t'']
\nonumber\\ &&-\,G_{\rm nl} \left[ v_2 - \frac{\beta_2}{ e}\int_0^\infty dt \frac{\pi T^2}{\sinh^2\pi Tt} K_{22}(t)\sin[ev_2t] \right]
\nonumber\\ &&
+\, \frac{1-\beta_1}{e} \int dt'dt''K_{12}(t''-t') \frac{{\cal D}(t',\bm{x}_1,\bm{x}_2)}{2e^2\nu_0R_1R_2}
\nonumber\\ &&\times\,
\frac{\pi T^2}{\sinh^2\pi Tt''}\sin[ev_2t''].
\label{current_final}
\end{eqnarray}
Here we have defined the response functions
\begin{eqnarray}
K_{ij}(t) = e^2\int \frac{d\omega}{2\pi}\frac{e^{-i\omega t}}{-i\omega +0}Z_{ij}(\omega),
\end{eqnarray}
which characterize the response of voltage fluctuations in the junction $i$ on the current noise of the junction $j$.
The corresponding impedance matrix $Z_{ij}(\omega)$ is defined in Appendix B, see Eq. (\ref{impedance}).
As before, here the voltage drops $v_1$ and $v_2$ are defined in Eq. (\ref{v1v2}).

Repeating now the same calculation in the elastic limit $L\ll L_{\rm in}$,
we obtain
\begin{eqnarray}
I_1 &=& G_{11}\bigg[v_1 - \frac{\beta_1}{e}\int_0^\infty dt \frac{\pi T^2}{\sinh^2\pi Tt} K_{11}(t)
\nonumber\\ &&\times\,
\left(\frac{r_2}{r}\sin[e(w_1-V_3)t]+\frac{r_1}{r}\sin[e(w_1-V_4)t]\right)\bigg]
\nonumber\\ &&
-\, G_{\rm nl} \bigg[ v_2 - \frac{\beta_2}{ e}\int_0^\infty dt \frac{\pi T^2}{\sinh^2\pi Tt} K_{22}(t)
\nonumber\\ &&\times\,
\left(\frac{r_2}{r}\sin[e(w_2-V_3)t]+\frac{r_1}{r}\sin[e(w_2-V_4)t]\right) \bigg]
\nonumber\\ &&
-\, \tilde G_{\rm nl}\frac{\beta_1}{e}\int_0^\infty dt \frac{\pi T^2}{\sinh^2\pi Tt} K_{11}(t) \sin[e(w_1-w_2)t]
\nonumber\\ &&
+\,\frac{\beta_1}{ e}\int dt'dt'' K_{12}(t''-t')
\frac{\pi T^2}{\sinh^2\pi Tt''}\frac{{\cal D}(t',\bm{x}_1,\bm{x}_2)}{2e^2\nu_0R_1R_2}
\nonumber\\ &&\times\,
\sin[(w_1-w_2)t'']
\nonumber\\ &&
+\, \frac{1-\beta_1}{e} \int dt'dt''K_{12}(t''-t') \frac{\pi T^2}{\sinh^2\pi Tt''}
\frac{{\cal D}(t',\bm{x}_1,\bm{x}_2)}{2e^2\nu_0R_1R_2}
\nonumber\\ &&\times\,
\left( \frac{r_2}{r}\sin[e(w_2-V_3)t'']+\frac{r_1}{r}\sin[e(w_2-V_4)t''] \right),
\label{current_elastic}
\end{eqnarray}
where the lead potentials $w_1,w_2$ are defined in Eq. (\ref{w1w2}).

Eqs. (\ref{current_final}), (\ref{current_elastic}) represent the central results of this paper
which fully determines the leading Coulomb corrections to the conductance matrix of our structure
in both relevant limits of strong and weak inelastic relaxation.
These results also allow to demonstrate a close relation between shot noise and interaction effects,
which is now extended to include non-local electron transport.
For example, the first line of Eq. (\ref{current_final}) describes the standard -- "local" --
Coulomb anomaly caused by charging effects and related to local shot noise \cite{GZ2001,yeyati}.
The next three lines in Eq. (\ref{current_final}) contain terms depending on the voltage difference $v_1-v_2$ and describing non-local effects.
Their origin can be traced
back to the corresponding contribution to the shot noise in the first junction, cf. the second line
in Eq. (\ref{S11_final}). Finally, the contribution in the last three lines in Eq. (\ref{current_final})
depends only on the voltage $v_2$ and emerges from the last term of Eq. (\ref{current_dc}) $\propto \langle\eta_2\rangle$.
In the same way one can establish the correspondence between various terms in the expressions for the current (\ref{current_elastic})
and noise (\ref{S11_elastic}), (\ref{S12_elastic}) in the elastic limit.
Perhaps we should also add that the above results remain applicable to a much
broader class of systems than that depicted in Fig. \ref{fig_two_junctions}.
E.g., the wire may be replaced by a metallic lead of any shape, and ultimately
all geometry specific details can be absorbed in few elements of the conductance matrix.

It is interesting to compare the results (\ref{current_final},\ref{current_elastic})
with the predictions of the $P(E)-$theory (\ref{PofE}).
Employing the usual definition of the $P(E)$ function\cite{Ingold} 
\begin{eqnarray}
P(E)&=&\int \frac{dt}{2\pi}\, e^{iEt + J_{11}(t)},
\label{PofE2}\\
J_{11}(t)&=& e^2\int\frac{d\omega}{2\pi}\,{\rm Re}\,\big[Z_{11}(\omega)\big]
\frac{[\cos\omega t-1]\coth\frac{\omega}{2T}+i\sin\omega t}{\omega},
\nonumber 
\end{eqnarray}
and combining it with the solution of the kinetic equation (\ref{kinetic_0})
one can evaluate the current (\ref{PofE}) in the limit of low resistances of the leads $h/e^2r_{j}\ll 1$.
Comparing the result with the Eqs. (\ref{current_final},\ref{current_elastic})
in the tunnel limit $\beta_1=\beta_2=1$, one observes that the $P(E)$ approach
reproduces the contributions containing the local response functions $K_{11}(t),K_{22}(t)$,
while the corrections $\propto K_{12}(t)$ are missing. One can further 
verify that the latter corrections originate from the cross-correlation of the
junction shot noises which are ignored in the formula (\ref{PofE2}).

\begin{figure}
\includegraphics[width=7cm]{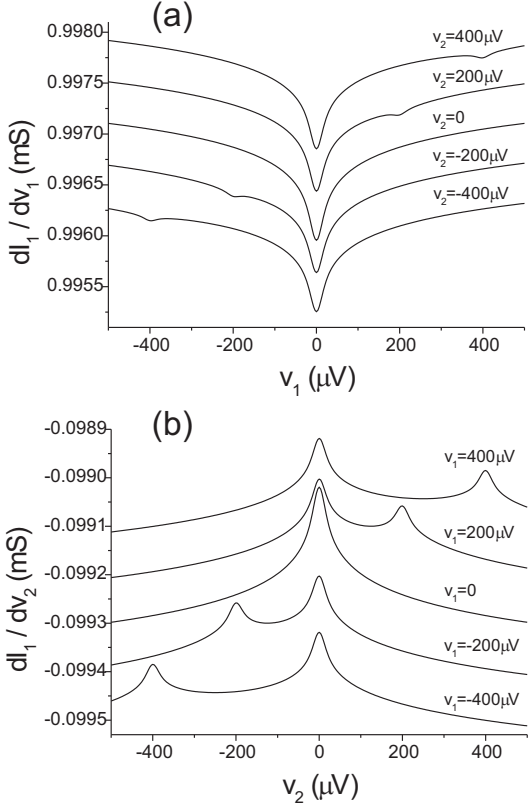}
\caption{Local (a) and non-local (b) differential conductances evaluated in the limit $|x_1-x_2|\ll L_{\rm in} \ll L$, Eqs. (\ref{di1dv1}) and (\ref{di1dv2}) respectively.
The system parameters are: $T=50$ mK, $\tau_0 = 1$ ns, $R_{S1}=3\Omega$, $R_{S2}=5\Omega$, $\beta_1=\beta_2=1$,
$G_{11}=1$ mS, $G_{\rm nl}=0.1$ mS.
The curves at $V_2=0$ in the top panel and at $V_1=0$ in the bottom panel are shown in real scale, other curves are shifted vertically for clarity. Local differential conductance $\partial I_1/\partial v_1$ exhibits a small dip at $v_1=v_2$. Non-local conductance $\partial I_1/\partial v_2$ shows a much more pronounced peak at $v_2=v_1$. }
\label{fig_didv1}
\end{figure}

In order to further specify our results it is necessary to make certain assumptions
about the form of the kernels $K_{ij}(t)$. For typical experimental setups and at sufficiently low voltages and temperature it is reasonable to adopt the following approximation for the elements of the admittance matrix of the environment $Y_{ij}(\omega)$ (see Eq. (\ref{admittance}) for their precise definition):
$Y_{11}(\omega)=1/R_{S1}$, $Y_{22}=1/R_{S2}$, $Y_{12}=Y_{21}=0$, where $R_{S1}$ and $R_{S2}$ are
effective shunt resistances. These resistances can roughly be estimated as
\begin{eqnarray}
R_{S1}=\bar r_1+{r_1r_2}/{(r_1+r_2)},\;\; R_{S2}=\bar r_2+{r_1r_2}/{(r_1+r_2)}.
\end{eqnarray}
In practice, the shunt resistances may deviate from these simple estimates due to impedance dispersion in metallic wires at high frequencies \cite{horizon}.
Further assuming that $G_{\rm nl}$ is small as compared to $Y_{11},Y_{22}$ one finds
$K_{12}(t)=K_{21}(t)=0$ and
\begin{eqnarray}
K_{11}(t) = e^2 R_{S1}\left( 1-e^{-t/\tau_0}\right),\;
K_{22}(t) = e^2 R_{S2}\left( 1-e^{-t/\tau_0}\right),
\nonumber
\end{eqnarray}
where $\tau_0 \sim R_{S1}C_1\sim R_{S2}C_2$ is the charge relaxation time which for simplicity
is taken equal for both junctions. This simplification is by no means restrictive since in our final result $\tau_0$ appears only under the logarithm as an effective cutoff parameter.
Under these conditions the current in the limit $L\gg L_{\rm in}$ (\ref{current_final}) can be evaluated analytically and takes the form
\begin{eqnarray}
I_1 &=& G_{11}\left[ v_1 - \frac{4\pi \beta_1 T }{eg_1} F_I(v_1)\right]
-G_{\rm nl}\left[ v_2 - \frac{4\pi \beta_2 T}{eg_2} F_I(v_2)\right]
\nonumber\\ &&
-\, \tilde G_{\rm nl} \frac{4\pi\beta_1 T}{eg_1} F_I(v_1-v_2),
\end{eqnarray}
where we defined the dimensionless conductances of the environment $g_1=2\pi/e^2R_{S1}$, $g_2=2\pi/e^2R_{S2}$ and
the dimensionless function
\begin{eqnarray}
F_I(v)&=&{\rm Im}\bigg[
\left(\frac{1}{2\pi T\tau_0}+i\frac{ev}{2\pi T}\right)\Psi\left(1+\frac{1}{2\pi T\tau_0}+i\frac{ev}{2\pi T}\right)
\nonumber\\ &&
-\,i\frac{ev}{2\pi T}\Psi\left( 1+i\frac{ev}{2\pi T} \right) \bigg].
\end{eqnarray}
Here $\Psi(x)$ stands for the digamma function.
Both local and non-local differential conductances read
\begin{eqnarray}
\frac{\partial I_1}{\partial v_1} = G_{11}\left[1-\frac{2\beta_1}{g_1} F(v_1)\right]
- \tilde G_{\rm nl} \frac{2\beta_1}{g_1} F(v_1-v_2),
\label{di1dv1}
\\
\frac{\partial I_1}{\partial v_2} = -G_{\rm nl}\left[1-\frac{2\beta_2}{g_2} F(v_2)\right]
+ \tilde G_{\rm nl} \frac{2\beta_1}{g_1} F(v_1-v_2),
\label{di1dv2}
\end{eqnarray}
where we  introduced another function
\begin{eqnarray}
F(v)&=&{\rm Re}\bigg[ \Psi\left(1+\frac{1}{2\pi T\tau_0}+i\frac{ev}{2\pi T}\right)
- \Psi\left( 1+i\frac{ev}{2\pi T} \right)
\nonumber\\ &&
+\,\left(\frac{1}{2\pi T\tau_0}+i\frac{ev}{2\pi T}\right)\Psi'\left(1+\frac{1}{2\pi T\tau_0}+i\frac{ev}{2\pi T}\right)
\nonumber\\ &&
-\,i\frac{ev}{2\pi T}\Psi'\left( 1+i\frac{ev}{2\pi T} \right) \bigg].
\label{f}
\end{eqnarray}

\begin{figure}
\begin{center}
\includegraphics[width=7cm]{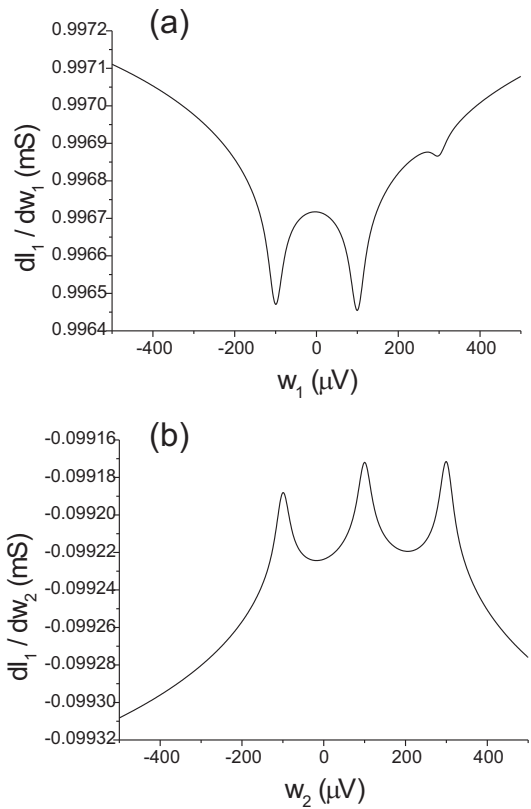}
\end{center}
\caption{Local (a)  and
non-local (b) differential conductances evaluated in the elastic limit $L\ll L_{\rm in}$, Eqs. (\ref{di1dw1}) and (\ref{di1dw2}) respectively.
The system parameters are the same as in Fig. 2. The voltage values
are:  $V_3=-100$ $\mu$V, $V_4=100$ $\mu$V,
$V_2=300$ $\mu$V in panel (a) and $V_1=300$ $\mu$V in panel (b). }
\label{fig_didv2}
\end{figure}

In the elastic limit $L\ll L_{\rm in}$ we find
\begin{eqnarray}
I_1 = G_{11}\bigg[ v_1 - \frac{4\pi \beta_1  T }{eg_1}\left( \frac{r_2}{r}F_I(w_1-V_3) + \frac{r_1}{r}F_I(w_1-V_4) \right)\bigg]
\nonumber\\
-\,G_{\rm nl}\bigg[ v_2 - \frac{4\pi \beta_2  T }{eg_2}\left( \frac{r_2}{r}F_I(w_2-V_3)+\frac{r_1}{r}F_I(w_2-V_4) \right) \bigg]
\nonumber\\
-\,  \tilde G_{\rm nl} \frac{4\pi\beta_1 T}{eg_1} F_I(w_1-w_2).\hspace{4.3cm}
\nonumber
\end{eqnarray}
Accordingly, local and non-local differential conductances acquire the form
\begin{eqnarray}
\frac{\partial I_1}{\partial w_1} &=& G_{11}\left[1-\frac{2\beta_1}{g_1} \left(\frac{r_2}{r}F(w_1-V_3)+\frac{r_1}{r}F(w_1-V_4)\right)\right]
\nonumber\\ &&
- \tilde G_{\rm nl} \frac{2\beta_1}{g_1} F(v_1-v_2),
\label{di1dw1}
\\
\frac{\partial I_1}{\partial w_2} &=& -G_{\rm nl}\left[1-\frac{2\beta_2}{g_2} \left(\frac{r_2}{r}F(w_2-V_3)+\frac{r_1}{r}F(w_2-V_4)\right) \right]
\nonumber\\ &&
+ \tilde G_{\rm nl} \frac{2\beta_1}{g_1} F(w_1-w_2).
\label{di1dw2}
\end{eqnarray}

Local differential conductance (\ref{di1dv1}) of a long wire with $L\gg L_{\rm in}$ is plotted in Fig. \ref{fig_didv1}a.
For a chosen set of parameters it is weakly affected by the second junction, although a small dip at $v_1=v_2$ is observed.
In contrast, non-local differential conductance $\partial I_1/\partial v_2$ is very sensitive to $v_1$ and has two peaks centered, respectively, at $v_2=0$ and $v_2=v_1$, see Fig. \ref{fig_didv1}b.
Fig. \ref{fig_didv2}a shows local differential conductance of a short wire, $L\ll L_{\rm in}$, in which the electron distribution function does not relax. We observe that the conductance $\partial I_1/\partial w_1$ given by Eq. (\ref{di1dw1})
has three dips centered respectively at $w_1=w_2,V_3,V_4$. Likewise,  non-local conductance
$\partial I_1/\partial w_2$ defined in Eq. (\ref{di1dw2}) shows peaks at $w_2=w_1,V_3,V_4$, see Fig. \ref{fig_didv2}b.
Comparing Figs. \ref{fig_didv1} and \ref{fig_didv2} we observe that
the dip in $\partial I_1/\partial v_1$ (the peak in $\partial I_1/\partial v_2$) occuring for strong inelastic electron relaxation splits into two dips (peaks) in the weak relaxation limit.

\section{Summary}
\label{sec:summary}

Let us briefly summarize our key observations.

We have demonstrated that Coulomb blockade corrections to both local
and non-local conductances in metallic conductors may change significantly provided the electron distribution function in at least one of the leads is driven out of equilibrium. Provided the conductor length $L$ is  shorter than the inelastic relaxation length $L_{\rm in}$, at low temperatures and under non-zero voltage bias the electron distribution
function acquires a characteristic double step form and the Coulomb dip in the differential conductance splits into two dips. This effect disappears provided inelastic relaxation becomes strong $L_{\rm in} \ll L$.

If two leads are attached to a metallic wire as it is shown in Fig. 1, the electron distribution function in the vicinity of one junction may also be driven out of equilibrium provided electrons are injected through the second junction and do not relax their energies at
distances shorter than the distance between these two junctions.
In this situation additional Coulomb dip in the differential conductance appears.

The latter configuration with two junctions also allows to study Coulomb blockade of non-local electron transport in the presence of non-equilibrium. It turns out that in this case an interplay between Coulomb and non-equilibrium effects yields more pronounced peaks
in the non-local differential conductance, see Figs. \ref{fig_didv1}b, \ref{fig_didv2}b. This observation indicates that experiments on non-local electron transport in the presence of Coulomb effects can be
conveniently used to test inelastic electron relaxation in metallic conductors at low temperatures,
as it has already been demonstrated, e.g., in experiments \cite{Anthore,Huard}.

The analysis developed here applies in the weak Coulomb blockade regime
implying that either the resistances
of metallic leads should be much smaller than the quantum resistance unit $h/e^2$, or the temperature should exceed charging energies of the barriers. In this regime there exists a transparent relation between shot noise and interaction effects in the electron transport \cite{GZ2001,yeyati}. Here we extended this fundamental relation to
the non-local case, demonstrating that negative cross-correlations in shot noise are directly linked to Coulomb suppression of non-local conductance. This is in contrast to NSN structures where Coulomb anti-blockade of non-local conductance may occur being related to positive cross-correlations in shot noise induced by crossed Andreev reflection.

\appendix

\section{Kinetic equation in the tunnel limit}
\label{sec:derivation}

Let us briefly discuss the main steps of our derivation of the kinetic equation (\ref{kinetic}). 

We start by defining the electron Keldysh Green function 
\begin{eqnarray}
G^K = \frac{\left\langle \hat\chi_\uparrow\left(t_1,{\bm x}_1 \right)
\hat\chi_\uparrow^\dagger\left(t_2,{\bm x}_2 \right)
- \hat\chi_\uparrow^\dagger\left(t_2,{\bm x}_2 \right)\hat\chi_\uparrow\left(t_1,{\bm x}_1 \right) \right\rangle}{2\pi\nu_0}
\label{GK}
\end{eqnarray}
This function obeys the equation
\begin{eqnarray}
\bigg( \frac{\partial}{\partial t_1} + \frac{\partial}{\partial t_2}
+i\frac{\nabla^2_{\bm{x}_2}-\nabla^2_{\bm{x}_1}}{2m} +iU(\bm{x}_1)-iU(\bm{x}_2)
\nonumber\\
+i\dot\Phi(t_1,\bm{x})-i\dot\Phi(t_2,\bm{x}_2)\bigg)G^K =
\nonumber\\
-\,i\frac{t^*_1(\bm{x})e^{-i\phi_1(t_1)}}{2\pi\nu_0}
\left\langle \hat\psi_{1,\uparrow}\left(X_1 \right)\hat\chi_\uparrow^\dagger\left(X_2 \right)
- \hat\chi_\uparrow^\dagger\left(X_2 \right)\hat\psi_{1,\uparrow}\left(X_1 \right) \right\rangle
\nonumber\\
+\,i\frac{t_1(\bm{x})e^{i\phi_1(t_2)}}{2\pi\nu_0}
\left\langle \hat\chi_\uparrow\left(X_1 \right)\hat\psi_{1,\uparrow}^\dagger\left(X_2 \right)
- \hat\psi_{1,\uparrow}^\dagger\left(X_2 \right)\hat\chi_\uparrow\left(X_1 \right) \right\rangle
\nonumber\\
-\,i\frac{t^*_2(\bm{x})e^{-i\phi_2(t_1)}}{2\pi\nu_0}
\left\langle \hat\psi_{2,\uparrow}\left(X_1 \right)\hat\chi_\uparrow^\dagger\left(X_2 \right)
- \hat\chi_\uparrow^\dagger\left(X_2 \right)\hat\psi_{2,\uparrow}\left(X_1 \right) \right\rangle
\nonumber\\
+\,i\frac{t_2(\bm{x})e^{i\phi_2(t_2)}}{2\pi\nu_0}
\left\langle \hat\chi_\uparrow\left(X_1 \right)\hat\psi_{2,\uparrow}^\dagger\left(X_2 \right)
- \hat\psi_{2,\uparrow}^\dagger\left(X_2 \right)\hat\chi_\uparrow\left(X_1 \right) \right\rangle.
\nonumber\\
\label{kinetic_GF}
\end{eqnarray}
Here we have defined the four-dimensional vectors $X_j=(t_j,\bm{x}_j)$.

Below we will stick to the diffusive limit in which case the electron distribution function remains isotropic. Then applying the standard quasiclassical technique \cite{Rammer},
we equalize the coordinates, $\bm{x}_1=\bm{x}_2=\bm{x}$, and make the replacement
\begin{eqnarray}
i\frac{\nabla^2_{\bm{x}_2}-\nabla^2_{\bm{x}_1}}{2m} +iU(\bm{x}_1)-iU(\bm{x}_2) \to -D\nabla^2_{\bm x}.
\label{repl}
\end{eqnarray}

We further note that the operators $\hat\psi_{j,\uparrow}$ and $\hat\chi_\uparrow$
in the vicinity of the barriers are not independent. They are related to each other via the scattering
matrices of the barrier.  Consider for simplicity the tunneling limit $T^{(j)}_n\ll 1$ in which case the corresponding transmission amplitudes in Eq. (\ref{ttr}) read
\begin{eqnarray}
\tau^{(j)}_n = -2\pi i \sqrt{\nu_j\nu_0} t_n^{(j)}.
\label{tunnel_amp}
\end{eqnarray}
Then we obtain
\begin{eqnarray}
\hat\chi_\uparrow(X_n) &=& \hat\chi_\uparrow^{\rm in}(X_n) + \sum_n\sum_{j=1,2}\tau_n^{(j)} e^{-i\phi_j(t_n)}\hat\psi_{j,\uparrow}^{\rm in}(X_n),
\nonumber\\
\hat\psi_{j,\uparrow}(X_n) &=& \hat\psi_{j,\uparrow}^{\rm in}(X_n) + \sum_n\tau_n^{(j)} e^{i\phi_j(t_n)}\hat\chi_{\uparrow}^{\rm in}(X_n),
\nonumber\\
\hat\chi_\uparrow^\dagger(X_n) &=& \hat\chi_\uparrow^{{\rm in}\dagger}(X_n)
+\sum_n\sum_{j=1,2}\left(\tau_n^{(j)}\right)^* e^{i\phi_j(t_n)}\hat\psi_{j,\uparrow}^{{\rm in}\dagger}(X_n),
\nonumber\\
\hat\psi_{j,\uparrow}^\dagger(X_n) &=& \hat\psi_{j,\uparrow}^{{\rm in}\dagger}(X_n)
+ \sum_n\left(\tau_n^{(j)}\right)^* e^{-i\phi_j(t_n)}\hat\chi_{\uparrow}^{{\rm in}\dagger}(X_n).
\label{relation}
\end{eqnarray}
Here the superscript $^{\rm in}$ labels incoming waves unaffected by the barriers.
In the tunneling limit considered here it suffices to identify the "incoming" operators with the full ones. Substituting the above expressions into Eq. (\ref{kinetic_GF}), performing the replacement (\ref{repl}) and setting $|t_j(\bm{x})|^2\propto \delta(\bm{x}-\bm{x}_j)$, after adding
the phenomenological term describing inelastic relaxation we arrive at Eq. (\ref{kinetic}) for the function $G(t_1,t_2,\bm{x})=G(t_1,t_2,\bm{x},\bm{x})$
without noise terms. The pre-factors in front of the terms on the right hand side of Eq. (\ref{kinetic})
are fixed by the requirement that in the absence of interactions the currents across the barriers have the standard Ohmic form $I_j=v_j/R_j$.

The noise terms may be derived if one employs Eq. (\ref{kinetic_GF}) for non-averaged operator
Green function
\begin{eqnarray}
\hat G^K = \frac{\hat\chi_\uparrow\left(t_1,{\bm x}_1 \right)\hat\chi_\uparrow^\dagger\left(t_2,{\bm x}_2 \right)
-\ \hat\chi_\uparrow^\dagger\left(t_2,{\bm x}_2 \right)\hat\chi_\uparrow\left(t_1,{\bm x}_1 \right)}{2\pi\nu_0}.
\label{GKna}
\end{eqnarray}
The noise operator $\hat\eta_1$ is then defined as follows
\begin{eqnarray}
&& \hat\eta_1(t_1,t_2)\propto
\nonumber\\ &&
-\,i\frac{et^*_1(\bm{x})e^{-i\phi_1(t_1)}}{2\pi}
\left( \hat\psi_{1,\uparrow}\left(X_1 \right)\hat\chi_\uparrow^\dagger\left(X_2 \right)
- \hat\chi_\uparrow^\dagger\left(X_2 \right)\hat\psi_{1,\uparrow}\left(X_1 \right) \right)
\nonumber\\ &&
+\,i\frac{et_1(\bm{x})e^{i\phi_1(t_2)}}{2\pi}
\left( \hat\chi_\uparrow\left(X_1 \right)\hat\psi_{1,\uparrow}^\dagger\left(X_2 \right)
- \hat\psi_{1,\uparrow}^\dagger\left(X_2 \right)\hat\chi_\uparrow\left(X_1 \right) \right)
\nonumber\\ &&
+\,i\frac{et^*_1(\bm{x})e^{-i\phi_1(t_1)}}{2\pi}
\left\langle \hat\psi_{1,\uparrow}\left(X_1 \right)\hat\chi_\uparrow^\dagger\left(X_2 \right)
- \hat\chi_\uparrow^\dagger\left(X_2 \right)\hat\psi_{1,\uparrow}\left(X_1 \right) \right\rangle
\nonumber\\ &&
-\,i\frac{et_1(\bm{x})e^{i\phi_1(t_2)}}{2\pi}
\left\langle \hat\chi_\uparrow\left(X_1 \right)\hat\psi_{1,\uparrow}^\dagger\left(X_2 \right)
- \hat\psi_{1,\uparrow}^\dagger\left(X_2 \right)\hat\chi_\uparrow\left(X_1 \right) \right\rangle.
\nonumber
\end{eqnarray}
Evaluating the symmetrized correlator of two such operators,
$$
\frac{1}{2}\left\langle \hat\eta_1(t_1,t_2)\hat\eta_1(t_3,t_4) + \hat\eta_1(t_3,t_4)\hat\eta_1(t_1,t_2)\right\rangle,
$$
one can verify that it coincides with the correlator (\ref{corr_eta})
in the tunneling limit $\beta_1=1$. The pre-factors in front of the noise terms
in the Eq. (\ref{kinetic}) are again determined by comparison with the noises of the junctions 
in the known non-interacting limit. 
The noise variable $\eta_2$ is defined analogously.
The operators $\hat G^K,\hat\eta_1,\hat\eta_2$ may be treated as classical fluctuating
functions in the spirit of the $\sigma-$model and path integral formulation \cite{KA}.

We finally note that the kinetic equation (\ref{kinetic}) can also be derived beyond the tunneling limit, i.e. for $T^{(j)}_n\sim 1$. However, in this general case the corresponding analysis turns rather complicated since the full scattering matrices of the barriers should be employed in Eq. (\ref{relation}).  Without going into such complicated algebra here we refer the reader to Ref. \onlinecite{array} where a general and rigorous derivation of the kinetic equation (\ref{kinetic}) has been carried out.

\section{Details of the solution of the kinetic equation}
\label{sec:solution}

In order to solve Eq. (\ref{kinetic}) we make use of the same procedure as in Sec. \ref{sec:noise}.
With the aid of Eq. (\ref{G_Keldysh}) the expression for the current (\ref{current_0}) can be rewritten as
\begin{eqnarray}
I_j(t) &=& \frac{\pi}{eR_j}\lim_{t'\to t}\left( G(t,t',\bm{x}_j)
- \frac{-iT e^{-i[\phi_1(t)-\phi_1(t')]}}{\sinh\pi T(t-t')} \right)
\nonumber\\ &&
+\, C_j\dot v_j+2\pi\eta_j(t,t),
\label{current_1}
\end{eqnarray}
where we added displacement currents recharging the capacitors $C_j$.
The solution of Eq. (\ref{kinetic}) takes the form
\begin{eqnarray}
&& G(t_1,t_2,\bm{x})=
\frac{-iT e^{-i[\Phi(t_1,\bm{x})-\Phi(t_2,\bm{x})]}}{\sinh\pi T(t_1-t_2)}
\int dt'  d^3\bm{x}'\,
\nonumber\\ &&
\frac{{\cal D}\left(\frac{t_1+t_2}{2}-t',\bm{x},\bm{x}'\right)}{\tau_{\rm in}} \,
\nonumber\\ &&
+ \, \frac{-iT e^{-i\left[\Phi(t_1,\bm{x})-\Phi(t_2,\bm{x})\right]}}{\sinh \pi T(t_1-t_2)}\int dt'
\nonumber\\ &&\times\,
\left[\frac{{\cal D}\left(\frac{t_1+t_2}{2}-t',\bm{x},\bm{x}_1\right)}{2e^2\nu_0R_1}
e^{-i\left[\varphi_1\left(t'+\frac{t_1-t_2}{2}\right)-\varphi_1\left(t'-\frac{t_1-t_2}{2}\right)\right]}
\right.
\nonumber\\ &&
\left.
+\, \frac{{\cal D}\left(\frac{t_1+t_2}{2}-t',\bm{x},\bm{x}_2\right)}{2e^2\nu_0R_2}
e^{-i\left[\varphi_2\left(t'+\frac{t_1-t_2}{2}\right)-\varphi_2\left(t'-\frac{t_1-t_2}{2}\right)\right]}
\right]
\nonumber\\ &&
-\, e^{-i\left[\Phi(t_1,\bm{x})-\Phi(t_2,\bm{x})\right]} \int dt'
\nonumber\\ &&\times\,
\bigg[
\frac{{\cal D}\left(\frac{t_1+t_2}{2}-t',\bm{x},\bm{x}_1\right)}{e\nu_0}
e^{i\left[\Phi\left(t'+\frac{t_1-t_2}{2},\bm{x}_1\right)-\Phi\left(t'-\frac{t_1-t_2}{2},\bm{x}_1\right)\right]}
\nonumber\\ &&\times\,
\eta_1\left(t'+\frac{t_1-t_2}{2},t'-\frac{t_1-t_2}{2}\right)
\nonumber\\ &&
+\,\frac{{\cal D}\left(\frac{t_1+t_2}{2}-t',\bm{x},\bm{x}_2\right)}{e\nu_0}
e^{i\left[\Phi\left(t'+\frac{t_1-t_2}{2},\bm{x}_2\right)-\Phi\left(t'-\frac{t_1-t_2}{2},\bm{x}_2\right)\right]}
\nonumber\\ &&\times\,
\eta_2\left(t'+\frac{t_1-t_2}{2},t'-\frac{t_1-t_2}{2}\right)
\bigg],
\label{G0}
\end{eqnarray}
where we defined $\varphi_j(t)=\phi_j(t)-\Phi(t,\bm{x}_j)=\int_{t_0}^t dt'\,ev_j(t')$.
Here we have already assumed that inelastic relaxation is strong, $L\ll L_{\rm in}$,
and that the the wire potential varies in space slowly enough, $e|V_3-V_4|\ll TL/L_{\rm in}$.

Combining Eqs. (\ref{current_1}) and (\ref{G0}) we evaluate the instantaneous current value in the first junction
\begin{eqnarray}
I_1(t) &=&  \int dt'
\bigg[\left(\frac{\delta(t-t')}{R_1}-\frac{{\cal D}_0\left(t-t',\bm{x}_1,\bm{x}_1\right)}{2e^2\nu_0R_1^2}\right)v_1(t')
\nonumber\\ &&
-\, \frac{{\cal D}_0\left(t-t',\bm{x}_1,\bm{x}_2\right)}{2e^2\nu_0R_1R_2}v_2(t')
\bigg]
+C_1\dot v_1+\delta I_1(t),
\label{current_2}
\end{eqnarray}
where the noise term $\delta I_1$ is defined in Eq. (\ref{xi1}) with the following replacement $\int dE\eta_j(t',E)\to 2\pi\eta_j(t',t')$.
Averaging the expression for the current (\ref{current_2}), (\ref{xi1}) over time we arrive at the following current-voltage characteristics
\begin{eqnarray}
I_1 &=& G_{11}v_1 - G_{\rm nl}v_2
\nonumber\\ &&
+\,2\pi R_1G_{11} \langle\eta_1(t,t)\rangle - 2\pi R_2G_{\rm nl} \langle\eta_2(t,t)\rangle.
\label{current_dc}
\end{eqnarray}
It is important to emphasize that here the average values $\langle\eta_j(t,t)\rangle$ differ from zero due to the presence of fluctuating phases which account for interaction effects.

In order to evaluate these averages it is convenient to split the time-dependent phases into regular and fluctuating parts,
\begin{eqnarray}
\varphi_j(t) = ev_j t +\delta\varphi_j(t),\;\;\; j=1,2,
\end{eqnarray}
where the potentials $v_j$ are defined in Eq. (\ref{v1v2}). In what follows we will assume that interaction effects remain sufficiently weak, which is the case provided either the resistances
of metallic wires are much smaller than the quantum resistance unit, $r_\alpha\ll h/e^2$, or the temperature is sufficiently high, $T > e^2/2C_j$. In either case phase fluctuations remain small, $\delta\varphi_j\ll 1$, and the average $\langle\eta_1(t,t)\rangle$ can be expressed in the form
\begin{eqnarray}
\langle\eta_1(t,t)\rangle = \int dt'\left\langle\frac{\delta\eta_1(t,t)}{\delta\varphi_1(t')}\delta\varphi_1(t')+
\frac{\delta\eta_1(t,t)}{\delta\varphi_2(t')}\delta\varphi_2(t')\right\rangle.
\label{av1}
\end{eqnarray}

Note that fluctuating phases $\delta\varphi_j(t)$, in turn, depend on the stochastic variables
$\eta_j(t)$. In order to establish this dependence we will make use of Fourier transformed Eq. (\ref{current_1}) which yields
\begin{eqnarray}
i_{1,\omega} = \left(-C_1\omega^2-i\omega G_{11}(\omega)\right) \frac{\delta\varphi_{1,\omega}}{e} +i\omega G_{\rm nl}(\omega) \frac{\delta\varphi_{2,\omega}}{e}
\nonumber \\
+\, 2\pi \left(-C_1\omega^2-i\omega G_{11}(\omega)\right) \eta_{1,\omega} + 2\pi i\omega G_{\rm nl}(\omega) \eta_{2,\omega},
\nonumber\\
i_{2,\omega} =  i\omega G_{\rm nl}(\omega) \frac{\delta\varphi_{1,\omega}}{e} + \left(-C_2\omega^2-i\omega G_{22}(\omega)\right) \frac{\delta\varphi_{2,\omega}}{e}
\nonumber\\
+\, 2\pi i\omega G_{\rm nl}(\omega) \eta_{1,\omega} + 2\pi \left(-C_2\omega^2-i\omega G_{22}(\omega)\right) \eta_{2,\omega}.
\nonumber
\end{eqnarray}
Here we introduced the Fourier transform of the fluctuating currents $i_{j,\omega}=\int dt e^{i\omega t}\left(I_j(t)-\langle I_j\rangle\right)$
and used the relation $\delta v_{j,\omega}=-i\omega \delta\varphi_{j,\omega}/e$.
The conductances $G_{11}(\omega)$, $G_{22}(\omega)$ and $G_{\rm nl}(\omega)$ are again defined in Eqs. (\ref{Gij}) where one should now substitute
$\tilde{\cal D}_0(0,\bm{x},\bm{x}')\to \tilde{\cal D}_0(\omega,\bm{x},\bm{x}')$,
i.e. these conductances are expressed via Fourier transformed diffusons at a frequency $\omega$.
From the equivalent circuit of Fig. 1b we can also define the fluctuating currents
\begin{eqnarray}
i_{i,\omega} = \sum_{j=1,2} i\omega Y_{ij}(\omega) \frac{\delta\varphi_{j,\omega}}{e},
\label{admittance}
\end{eqnarray}
where $Y_{ij}(\omega)$ is the admittance matrix of our structure. The off-diagonal
elements $Y_{12}(\omega)=Y_{21}(\omega)$ are responsible for cross-correlations between the junctions,
which may be caused, e.g., by capacitive coupling between the leads 1 and 2.
Excluding the currents $i_{\omega,j}$ from the above equations we obtain
\begin{eqnarray}
\delta\varphi_i(t)= -\frac{2\pi}{e}\sum_{j=1,2}\int dt' K_{ij}(t-t')\eta_j(t'),
\label{31}
\end{eqnarray}
where the kernels $K_{ij}(t)$ read
\begin{eqnarray}
K_{ij}(t) = e^2\int \frac{d\omega}{2\pi}\frac{e^{-i\omega t}}{-i\omega +0}Z_{ij}(\omega),
\end{eqnarray}
with $Z_{ij}(\omega)$ being an effective impedance matrix
\begin{eqnarray}
Z_{ij}(\omega)=
\left(\begin{array}{cc}
\frac{-i\omega C_2+ G_{22}(\omega) + Y_{22}(\omega)}{A(\omega)} & \frac{G_{\rm nl}(\omega)+Y_{12}(\omega)}{A(\omega)} \\
\frac{G_{\rm nl}(\omega)+Y_{21}(\omega)}{A(\omega)}  & \frac{-i\omega C_1+ G_{11}(\omega) + Y_{11}(\omega)}{A(\omega)} \\
\end{array}\right).
\label{impedance}
\end{eqnarray}
and
\begin{eqnarray}
A(\omega)&=&\big(-i\omega C_1\omega^2 + G_{11}(\omega)+Y_{11}(\omega)\big)
\nonumber\\ &&\times\,
\big(-i\omega C_2+G_{22}(\omega)+Y_{22}(\omega)\big)
\nonumber\\ &&
-\,\big(G_{\rm nl}(\omega)+Y_{12}(\omega)\big)^2.
\nonumber
\end{eqnarray}

Combining Eqs. (\ref{av1}) and (\ref{31}), we obtain
\begin{eqnarray}
\langle\eta_1(t,t)\rangle &=& -\frac{2\pi}{e}\sum_{j,k=1,2}\int dt' dt''\,
K_{jk}(t'-t'')
\nonumber\\ &&\times\,
\left\langle\frac{\delta\eta_1(t,t)}{\delta\varphi_j(t')}\eta_k(t'',t'')\right\rangle.
\label{av2}
\end{eqnarray}
Due to causality the variable $\eta_1(t)$ can only depend on the phases $\varphi_j(t')$ taken at earlier times (i.e. at $t'<t$), while the function $K_{ij}(t'-t'')$
differs from zero only for $t'>t''$. Hence, the variable $\eta_k(t'')$ is independent of $\varphi_j(t')$, and Eq. (\ref{av2}) can be rewritten in the form
\begin{eqnarray}
\langle\eta_1(t,t)\rangle &=& -\frac{2\pi}{e}\sum_{j=1,2}\int dt' dt''\,K_{j1}(t'-t'')
\nonumber\\ &&\times\,
\left.\frac{\delta}{\delta\varphi_j(t')}\left\langle \eta_1(t,t)\eta_1(t'',t'')\right\rangle
\right|_{\varphi_j = eV_jt}.
\label{av3}
\end{eqnarray}
Here the correlator $\left\langle \eta_1(t,t)\eta_1(t'',t'')\right\rangle$ is defined in Eq. (\ref{corr_eta}) with the function $G(t,t'',\bm{x}_1)$ set by Eq. (\ref{G0}) with omitted noise terms, i.e. with $\eta_{1,2}=0$.
The average value $\langle\eta_2\rangle$ is derived in exactly the same manner.

Now we are in a position to evaluate the functional derivative $\delta\left\langle \eta_1(t,t)\eta_1(t'',t'')\right\rangle/\delta\varphi_j(t')$
from Eq. (\ref{corr_eta}). After a straightforward but rather tedious calculation one
arrives at the result (\ref{current_final}).

\end{document}